\documentclass[natbib, twocolumns, a4paper]{aastex}

\usepackage{amsmath}                   
\usepackage{amssymb}                   
\usepackage{bm}                        
\usepackage[british]{babel}            
\usepackage{enumerate}                 
\usepackage{natbib}                    
\usepackage{hyperref}                  
\hypersetup{                           
	colorlinks=true, 
	citecolor=blue, 
	filecolor=blue,
	linkcolor=blue, 
	urlcolor=blue
	}
\usepackage{multirow}                  
\usepackage{graphicx}                  
\usepackage{subfigure}                 
\usepackage{mathrsfs}                  
\usepackage{bbding}

\newcommand{\be}{\begin{equation}}
\newcommand{\bs}{\begin{subequations}}
\newcommand{\cm}{{~\rm cm}}
\newcommand{\der}[2]{\frac{d\,{#1}}{d\,{#2}}}
\newcommand{\dpar}[2]{\frac{\partial {#1}}{\partial {#2}}}
\newcommand{\ee}{\end{equation}}
\newcommand{\ergs}{{~\rm erg\, s^{-1}}}
\newcommand{\es}{\end{subequations}}
\newcommand{\g}{{~\rm g}}

\newcommand{\keV}{{~\rm keV}}
\newcommand{\kms}{{~\rm km\, s^{-1}}}

\newcommand{\mdot}{{\dot{M}}}

\newcommand{\mean}[1]{{\langle{#1}\rangle}}
\newcommand{\mns}{{M_{\rm ns}}}
\newcommand{\mnsU}{{\left(\frac{M_{\rm ns}}{1.4~M_{\odot}}\right)}}
\newcommand{\msun}{{~M_{\odot}}}

\newcommand{\myr}{{~M_{\odot}\, \rm yr^{-1}}}

\newcommand{\rns}{{R_{\rm ns}}}
\newcommand{\rst}{{R_{\rm st}}}

\newcommand{\s}{{~\rm s}}

\shorttitle{Stochastic accretion}
\shortauthors{Pizzolato \& Sidoli}

\slugcomment{Draft version of \today}
\slugcomment{Astrophysical Journal, submitted}

\begin{document}

\title{Stochastic accretion and the variability of supergiant  fast X--ray transients}

\author{
        Fabio Pizzolato\altaffilmark{1}  
        \and
        Lara Sidoli\altaffilmark{1}  
       }

\altaffiltext{1}{
	INAF-IASF Milano, Via Bassini no. 15, I-20133 Milano,   Italy
}
	\email{\href{mailto:fabio@iasf-milano.inaf.it}{fabio@iasf-milano.inaf.it}}

	\email{\href{mailto:sidoli@iasf-milano.inaf.it}{sidoli@iasf-milano.inaf.it}}


\begin{abstract}

In this paper we consider the variability of the luminosity 
of a compact object (CO) powered by the accretion of an
extremely inhomogeneous (``clumpy'') stream of matter.
The accretion of a single clump results in an X-ray flare: we adopt a
simple model for the response of the CO to the accretion of a single clump, 
and  derive a stochastic differential equation (SDE) 
for the accretion powered luminosity $L(t)$.
We put the SDE in the equivalent form of  an  equation for the flares' luminosity distribution (FLD), and discuss its solution in the stationary case.

As a case study, we apply our formalism to the analysis of the FLDs
of Super-Giant Fast X-ray Transients (SFXTs),
a peculiar sub-class of  High Mass X--ray Binary Systems (HMXBs).
We compare our theoretical FLDs to the distributions observed in
the SFXTs $IGR~J16479-4514$, $IGR~J17544-2619$ and  $XTE~J1739-302$.

Despite its simplicity, our model fairly agrees with the 
observed distributions, and allows to predict some
properties of the stellar wind. Finally, we  discuss 
how our model may explain the difference between the 
broad FLD of SFXTs and the much narrower
distribution of persistent HMXBs.

\end{abstract}




\section{Introduction}
\label{s-intro}

In several astrophysical contexts the mass accretion 
onto a compact object (CO, a black hole, a neutron star a 
or white dwarf) cannot be considered as a continuous process. 
The accretion may be  discrete:
the mass does not flow as a continuous stream,
but consists of a volley of  clumps
whose accretion on the compact object results in 
a flaring activity observed in the X--ray/$\gamma$ ray spectral window.
Inhomogeneous flows occur in the
Polar  (a.k.a. AM~Herculis) cataclysmic variables 
(e.g. \citealp{Warner:1995a}),  or in the Supergiant Fast X--ray
Transients (\citealp{Sguera:2006}, \citealp{Negueruela:2005a, Zand:2005}). Understanding the properties of clumpy accretion 
on the super--massive black holes hosted at the core of galaxy clusters
is also relevant in the context of the ``cold feedback model'' 
(e.g. \citealp{Pizzolato:2005a}).

The luminosity $L(t)$ powered by the accretion  on a compact object
results from  the interplay of  two factors: the properties of the 
accreting stream and the response of the CO to their arrival.
The response process may be quite complex, but it is essentially deterministic.
On the other hand, if the accretion occurs from a population 
of clumps with random masses arriving  at random times, 
the accretion driving term is a stochastic process. The 
accretion-powered luminosity  $L(t)$ is an irregular function of
time, and it requires an adequate mathematical tool to be 
dealt with. This is provided by the theory of stochastic differential
equations (SDEs), that replaces the rules of
ordinary calculus to treat highly irregular functions (see e.g. \citealp{Oksendal:2003a, Gardiner:2009a}). 
 
In this paper we consider the variability of the X/$\gamma$ ray
luminosity of a CO powered by the accretion of a family of clumps 
with randomly distributed masses. 
We shall derive (Sec.~\ref{s-stacc}) a simple SDE for the luminosity $L(t)$, and will illustrate some of the subtleties involved
in handling SDEs.  A possible way to use this SDE is to compute a sample
path of $L(t)$, to be compared to an observed light curve. 
An equivalent approach
(presented in Sec.~\ref{s-gfpe}) consists in  associating to 
the SDE for $L(t)$ an integro-differential equation
(known as ``generalised Fokker-Planck equation'', GFPE) for the 
probability function $p(L, t)$ that the source has luminosity $L$
at the time $t$.  We shall discuss the properties of the distribution 
$p$, and how it mirrors the features of the accreting stream.
As a case study we apply our model to the super-giant Fast X--ray
Transients (SFXTs), an interesting class of High Mass X--ray binaries composed
by a neutron star (or a black hole) accreting  an extremely 
inhomogeneous wind blown by its massive companion. 
Our analysis of SFXTs is  presented
in Sec.~\ref{s-sfxt}.
We shall discuss our findings and in 
particular how our model helps to explain the  difference 
of the FLDs of SFXTs from those  of the much more common 
persistent X-ray sources.
We shall summarise in Sec.~\ref{s-summary}.


\section{A simple stochastic accretion model}
\label{s-stacc}

The analysis of the long term variability of the X--ray light curve
generated by the inhomogeneous accretion requires some modelling of the
response of the CO the the accretion process of a single clump.
To this purpose it is helpful to imagine the CO as 
a ``black box'' that emits X--rays in response to the 
accretion of mass from the surrounding environment. If $\mdot_{c}$ is the mass capture rate and $L$ is the luminosity produced by the CO,
we postulate the linear response
\be
\label{e-lx}
L(t) = \int_{\mathbb R} d t' \; W(\mathbf p, t-t') \: \mdot_{c}(t'),
\ee
where the function $W$ (dependent on the parameters $\mathbf p$)
describes the response of the CO.

The analytical shape of $W$ may be exceedingly complex, since it
must embody  a number of physical processes: tidal effects, interaction
of the accretion flow with the magnetic field of the CO, 
geometry and plasma instabilities of the accretion stream, radiation effects, 
and so on.
Since this paper aims to focus on the basic properties of the accretion process,
we steer clear of these  details and demand for $W$ just few very  basic properties;
i)~$W(t)$ is causal: it vanishes for $t<0$, i.e. 
before the capture of a clump, and ii)~it decays to zero for $t\gg0$, i.e. 
long after the clump has been captured.
The simplest function with these properties is 
\begin{align}
\label{e-window}
{W}(t) = 
\left\{
	\begin{aligned}
	&0  & t< 0 
	\\
	& \frac{G\, M}{\rst}\;\frac{ {\rm e}^{-t/\tau} }{\tau}    & t\geq0
	\end{aligned}
\right..
\end{align}
This response gives the accretion powered  luminosity of a unit mass
on a  CO of mass $M$,  down to a ``stopping radius'' $\rst$. If the
CO is able to accrete the flow down to its radius (or event horizon) $R$, 
then $\rst = R$. In some cases, e.g. a fast spinning magnetised
neutron  star, the so called  ``propeller'' effect 
(see e.g. \citealp{Lipunov:1987a, Illarionov:1975a, Davies:1979a, Davies:1981a})
prevents the stream to flow beyond the magnetosphere, and in this case
$\rst$ is the Alfv\'en radius $R_{A}$.
The time scale $\tau$ represents the time scale taken by the CO to
``process'' the accretion stream. For example, if the stream has
an appreciable angular momentum, it forms an accretion disc before falling
on the CO, and the mass accretion rate on the CO 
decays as a power law  on the viscous time scale
\be
\label{e-tvisc}
\tau\simeq
7.5\times 10^{5}~{\rm s} \;
\alpha^{-4/5} \:
\left(\frac{\mdot}{10^{-11}\myr}\right)^{-3/10}\;
\mnsU^{1/4}\;
\left(\frac{R_{\rm out}}{10^{10}\cm}\right)^{5/4},
\ee
(Eq.~5.63 in \citealp{Frank:2002a}), 
where  $R_{\rm out}$ is the  outer radius and
$\alpha\lesssim 1$ is the viscosity  parameter \citep{Shakura:1973a}.

It is instructive to consider Eqns.~\eqref{e-lx} and~\eqref{e-window} 
in the simple case  of the accretion a single clump of mass $m$, 
with $\mdot_{c}(t) = m\, \delta(t)$ (where $\delta$ is the Dirac delta function).
With our choice~\eqref{e-window} of the response $W$, the luminosity  is
zero at $t<0$, i.e. before the clumps arrives, and
\be
L = \frac{G\, M}{R_{\rm st}} \: \frac{m}{\tau}\;  {\rm e}^{-t/\tau}
\ee
for $t\geq 0$. The luminosity has a sharp peak, and then decays
exponentially with the e-folding time $\tau$.
Plugging the Ansatz~\eqref{e-window} into Eq.~\eqref{e-lx} we find 
\be
\label{e-sde-0}
\tau \: \der{L}{t} = \frac{G\, M}{R_{\rm st}}\; \mdot_{c} - L. 
\ee 
We introduce the mass accretion rate $\mdot$ 
accreted  down to the stopping radius $\rst$
\be
\label{e-lxmdot}
L\equiv \frac{G\, M}{\rst} \: \mdot,
\ee 
so Eq.~\eqref{e-sde-0} becomes
\be
\label{e-sde-1}
\tau \: \der{\mdot}{t} = \mdot_{c} - \mdot.
\ee
Since the function $\mdot_{c}(t)$ is random, the function $\mdot(t)$
is highly irregular and certainly not differentiable, so Eq.~\eqref{e-sde-1}
is to be  interpreted as a stochastic differential equation 
(see e.g. \citealp{Oksendal:2003a}), and not as an ordinary differential equation.

Thus far we have provided a simple model for the ``response'' of
the compact object to the accretion of a clump of matter.
We now turn our attention to  the driving stochastic  term $\mdot_{c}$,
which embodies both the arrival rate of the clumps and their mass
distribution. Neglecting the finite size of the clumps,
we model the clumps' arrival rate as a train of
delta pulses 
\be
\label{e-train}
\mdot_{c}(t) = \sum_{k=1}^{n(t)} m_{k} \: \delta(t - t_{k}),
\ee
where $m_{k}$ is the mass of the clump accreted at the time $t_{k}$, 
and $n(t)$ is a Poisson counting process,  described by the probability
\be
\label{e-poisson}
{\mathcal P}[n(t) = n] = \frac{ {\rm e}^{-\lambda\, t}\, (\lambda\,t)^{n}}{n!},
\ee 
where $\lambda$ is the clumps' arrival rate.
The  masses  $\{m_{k}\}$ are distributed  according to the 
probability distribution function  $\varphi(m)$; we assume 
that $\varphi $ is stationary, and
that the random variables $\{m_{k}\}$ and  $\{t_{k}\}$
are uncorrelated.
The statistical properties of $\mdot_{c}$ are readily derived:
\bs
\label{e-process}
\begin{align}
\label{e-mean}
 \mean{\mdot_{c}} &= \lambda \; \mean{m}
\\
\label{e-ac}
\mean{\mdot_{c}(t)\: \mdot_{c}(t')} &= \lambda \; \mean{m^{2}}\;
\delta(t-t').
\end{align}
\es
The properties~\eqref{e-process} show that the Poisson 
process~\eqref{e-train} is a white noise with non--zero mean.

\bigskip

From Eqns.~\eqref{e-sde-1}, \eqref{e-train} and~\eqref{e-process} it is 
possible to derive some properties of the random 
function $\mdot(t)$.  They not only provide a useful check against our  more elaborate  approach developed in the next sections, but they also
help to point out some subtle key properties of SDEs.

Integrating Eq.~\eqref{e-sde-1}, we find the expression
\be
\label{e-sde-2}
\tau \, d\mdot = \delta m(t) - \mdot(t) \, dt, 
\ee
where
\be
\delta m(t) = \int_{t}^{t + d\,t} dt' \mdot_{c}(t'). 
\ee
With the aid of Eqns.~\eqref{e-process} we derive
\bs
\label{e-12moments}
\begin{align}
\label{e-1moment}
& \mean{\delta m} = \int_{t}^{t+d t} dt' \mean{\mdot_{c}(t')}
= \lambda\ \mean{m} \,  dt 
\\
\label{e-2moment}
& \mean{ (\delta m)^{2}}  = 
\int_{t}^{t+d t} dt' 
\int _{t}^{t+ d t} dt''
\mean{ \mdot_{c}(t')\mdot_{c}(t'')} = \lambda\ \mean{m^{2}} \, dt.
\end{align}
\es
With the first of these expressions we find the 
mean of Eq.~\eqref{e-sde-2}, i.e.  the ordinary differential equation
\be
\tau \, \der{\mean{\mdot}}{t} = \lambda\, \mean{m} - \mean{\mdot}  
\ee
(the order of the mean and the derivative may be exchanged, 
see e.g. \citealp{Reif:1965a}).

For $t\gg \tau$ the mean mass accretion rate $\mean{\mdot}$ relaxes on the stationary value
\be
\label{e-mean}
\mean{\mdot} =\lambda\, \mean{m}. 
\ee
The evaluation of the second moment is trickier, on account of
the stochastic nature of the equation. We write
\be
\label{e-dmx}
\tau \, d(\mdot^{2}) = \tau \left[(\mdot + d\, \mdot)^{2} - \mdot^{2}\right] =
\tau\, \left[2 \mdot d\mdot + (d\mdot)^{2} \right], 
\ee
where in the expansion we must keep the second order term $(d\mdot)^{2}$
for reason that will be clear soon.
Plugging Eq.~\eqref{e-sde-2} into this expression we get
\be 
\label{e-dm2}
\tau \, d(\mdot^{2}) =
2\, \mdot\left( \delta m - \mdot(t)\,dt\right) + (\delta m)^{2} + O(dt),
\ee
and taking the mean by using  Eqns.~\eqref{e-12moments}
we find the ordinary differential equation
\be
\tau\, \der{\mean{\mdot^{2}}}{t}  = 
2 \lambda^{2} \mean{m}^{2} - 2\, \mean{\mdot^{2}} + \frac{\lambda}{\tau} 
\, \mean{m^{2}}. 
\ee
For $t\gg \tau$ the second moment $\mean{\mdot^{2}}$ relaxes on the stationary value
\be
\mean{\mdot^{2}} = \lambda^{2} \mean{m}^{2} + \frac{\lambda}{2\, \tau} \mean{m^{2}},
\ee
corresponding to the variance
\be
\label{e-var}
\sigma_{\mdot}^{2} = \mean{\mdot^{2}} - \mean{\mdot}^{2} = 
\frac{\lambda}{2\, \tau} \mean{m^{2}}.
\ee
Some remarks are in order. 
First, the stochastic nature of the function $\mdot_{c}$ makes
$\mdot(t)$ an extremely irregular function of time. On account of
such irregular behaviour, none of the rules of ordinary calculus
are applicable to $\mdot(t)$. One consequence is that the second order differential $(d\mdot)^{2}$ is proportional to $dt$, and therefore it 
cannot be neglected in the formal manipulations, 
as we have done in Eq.~\eqref{e-dmx}. 
In order to work with such irregular  functions a new
kind  of differential calculus (known as ``It\={o} calculus'') has    
been developed  as a  corner stone of the theory of
stochastic differential equations (SDEs, see e.g. \citealp{Oksendal:2003a}).
Some level of knowledge of SDEs is therefore
essential to master the modelling of random processes like that 
studied in this paper. 

The moments $\mean{\mdot^{n}}$ exist 
only if the correspondent moments $\mean{m^{n}}$ of the clumps' distribution 
are  defined.  This is not always the case, e.g. when the
distribution $\varphi$ has a fat tail. In order to deal with such  
possibility, it is necessary to extend the simple approach presented in this 
section. We shall explain how in Sec.~\ref{s-dsde}.  

The mean accretion rate~\eqref{e-mean} depends on the clumps' 
arrival rate $\lambda$, but not from the relaxation time $\tau$, which is
an intrinsic property of the accretor. 
The variance~\eqref{e-var} features the relaxation time $\tau$  with an inverse power: a  long relaxation time $\tau$ corresponds to  a narrow dispersion 
of the observed values of $\mdot$  around the mean. Indeed, if  the 
relaxation  time $\tau$ is long, several clumps may be accreted before the
accretor is able to respond, and an observer cannot
distinguish between several elementary accretion processes, which 
are perceived as a single accretion  of a large clump. Clearly, this
has the effect of reducing the dispersion of the observed distribution.

\subsection{Direct SDE vs Fokker-Planck approach}
\label{s-dsde}

In principle, Eq.~\eqref{e-sde-1} may be solved
numerically. Several computing techniques have been developed
to tackle the numerical solution of SDEs 
(see e.g.~\citealp{Kloeden:1999a}).
A sample of the population of the clumps' masses  and arrival times are
extracted from the distributions $\varphi$ and ${\mathcal P}[n(t)]$ 
(defined by Eq.~\eqref{e-poisson}).
A sample path of the process $\mdot(t)$ is computed numerically,
and  $\mdot(t)$ is  converted to the  accretion luminosity $L(t)$ 
via Eq.~\eqref{e-lxmdot}. A sample path  $L_{1}(t)$ may be very different from
another path $L_{2}(t)$, since the clumps' masses and arrival times
extracted from the distributions $\varphi$ and ${\mathcal P}[n(t)]$ used to 
compute $L_{1}(t)$ numerically 
may be very different from those extracted to compute $L_{2}(t)$.
For this reason,  it is  meaningless to 
compare a {\em single} sample path of $L(t)$ with a real
light curve.
Any comparison of a stochastic model with the observations will
involve some statistics on a sample of the computed paths and 
the light curves.

We shall adopt an equivalent approach to achieve the goal of 
comparing a theoretical stochastic light curve with  a real
light curve. It is possible to associate to Eq.~\eqref{e-sde-1}
a new equation (called a ``generalised Fokker--Planck equation'', GFPE) 
for the probability $p(L,t)$ that the source has luminosity $L$ at the time 
$t$.  The distribution $p$ is directly comparable to the 
histogram of the flares' luminosities observed in a real source
over a long time span. 

The approach based on the luminosity distribution $p(L,t)$
allows to deal quite straightforwardly with a complication we have overlooked thus far.
The  luminosity distribution is defined as a function of
the  X--ray luminosity $L$, not of the mass accretion rate $\mdot$.
The general relation between the luminosity distribution
$p(L, t)$ and the mass accretion rate distribution $P(\mdot,t)$ is
\be
\label{e-pw}
p(L, t) =   \left|\der{\mdot}{L}\right|\:  P(\mdot, t)  
\ee
If all the mass can reach the surface
of the neutron star (located at the radius $R$), 
then $R_{\rm st} = R$.  
In this case the distributions $p(L)$ and $P(\mdot)$  
are simply proportional to each other:
\be
\label{e-lum-direct}
p(L, t) = \frac{R}{G\, M} \; P(\mdot, t).
\ee
In some systems, however, the accreting stream cannot
reach the surface: this is typically the case
of a magnetised fast spinning neutron star in which the matter 
is captured by the
star's gravitational field, but is prevented from reaching its surface by 
a ``propeller'' barrier. 
At a very basic level, the ``propeller'' effect may be described as follows.
The matter inside the magnetosphere (located at the Alfv\'en radius
$R_{A}$)  is forced to rotate with the same
angular velocity $\omega_{\rm ns}$ as the neutron star. 
On the other hand, the angular velocity of the mass stream 
outside the magnetosphere ($r>R_{A}$) is  approximately keplerian, 
$\omega\simeq \omega_{K}(r)$.  If 
\be
\label{e-propeller}
\omega_{\rm ns} >\omega_{K}(R_{A})
\ee 
no accretion is possible, and the
matter is stopped at the Alfv\'en radius. 
The radius $R_{A}$ depends on  the geometry of
the accretion flow. If NS accretes a clumpy wind in a detached 
binary system  the 
accretion is approximately spherical, and there are two possible
regimes. The NS crosses a wind blown by the companion star with 
a velocity $v_{w}$. 
This wind is focused by the NS's own gravitational field within
the capture radius
\be
\label{e-rcapt}
R_{G} =  \frac{2\, G \,\mns}{v_{w}^{2}}
=
3.74\times 10^{10}\cm \; \mnsU \; \left(\frac{v_{w}}{10^{3}\kms}\right)^{-2}.
\ee 
If the magnetic field of the NS is ``weak'', then the wind is first focused
by the NS's gravitational field, and only then it is affected by the NS'
magnetic force.
In this case the correct expression of the Alfv\'en radius is
(e.g. \citealp{Lipunov:1987a})
\be
\label{e-ra-1}
R_{A} = \left(\frac{\mu^{4}}{2\,G\, \mns\,\mdot^{2}}\right)^{1/7} 
\qquad \text{if} \qquad R_{A}<R_{G},
\ee
where $\mu$ is the magnetic moment of the NS.
On the other hand, if the magnetic field is strong, then the wind
feels NS' magnetic  field before its gravity, and in this regime
\be
\label{e-ra-2}
R_{A} = \left(\frac{4\, \mu^{2} G^{2}\,\mns^{2}}{\mdot\, v_{w}^{5}}\right)^{1/6}
\qquad \text{if} \qquad R_{G} < R_{A}.
\ee

The critical mass accretion rate $\mdot_{\omega}$
at which the propeller barrier sets in can be derived from Eq.~\eqref{e-propeller}
with $R_{A}$ given by either Eq.~\eqref{e-ra-1} or~Eq.~\eqref{e-ra-2}.
If $\mdot<\mdot_{\omega}$ the ram pressure exerted by the 
flow is unable to overcome the magnetic pressure at the magnetospheric
radius, and the flow is stopped at the Alfv\'en radius $R_{A}$.
The stopping radius is then
\begin{align}
\rst = 
	\left \{
	\begin{aligned}
	& R                                 & \mdot > \mdot_{\omega}
	\\
	& R_{A}                             & \mdot < \mdot_{\omega}  
	\end{aligned}
	\right..
\end{align}
Plugging this stepwise function $\rst(\mdot)$ into
Eq.~\eqref{e-pw} we find
\begin{align}
\label{e-pl}
p(L,t) = 
	\left \{
	\begin{aligned}
	& \frac{R_{A}}{G\,M}\; P(\mdot, t)  & L < L_{1}
	\\
	& 0                            & L_{1} < L <  L_{\omega}
	\\
	& \frac{R}{G\, M}\; P(\mdot, t) & L> L_{\omega}  
	\end{aligned}
	\right.,
\end{align}
where $L_{1}=G\,M\mdot_{\omega}/R_{A}$ and
$ L_{\omega} = G\, M\,\mdot_{\omega} / R$. 

For a typical parameters of a NS ($\mu\simeq 10^{30}~\rm G$)
and of a stellar wind ($v_{w}\simeq 10^{3}\kms$)
the luminosity $L_{1}$ is below $\sim 10^{32}\ergs$, which
is barely observable.  Therefore, we neglect the
tail below $L_{1}$ and approximate  Eq.~\eqref{e-pl} with
\begin{align}
\label{e-lum-propeller}
p(L,t) \simeq
	\left \{
	\begin{aligned}	
	& 0                             & \mdot < \mdot_{\omega}
	\\
	& \frac{R}{G\, M}\; P(\mdot, t) & \mdot > \mdot_{\omega}  
	\end{aligned}
	\right..
\end{align}
In presence of an accretion barrier the luminosity distribution  
$p(L)$ has  a sharp lower cut--off at the luminosity $L_{\omega}$.

Whether an accretion barrier is present or not, $p$ must be
computed from the  mass accretion distribution
$P(\mdot, t)$.  If no accretion barrier is present, $p$  is
given by $P(\mdot, t)$ and Eq~\eqref{e-lum-direct},  
else it is computed from $P(\mdot, t)$ and Eq.~\eqref{e-lum-propeller}.

In the next section we derive and discuss  the GFPE for 
$P(\mdot, t)$ associated to
the stochastic differential  equation~\eqref{e-sde-1}.


\section{The generalized Fokker--Planck equation}
\label{s-gfpe}

The  method to derive the GFPE  from a stochastic differential equation
is described by \citet{Denisov:2009a}. 
We  leave all the technical details of the derivation in 
Appendix~\ref{a-gfpe}, and here we reproduce only the results
derived there. The GFPE associated to Eq.~\eqref{e-sde-1}  reads
\bs
\label{e-bvp}
\be
\label{e-gfpe}
\tau\,\dpar{P}{t} = \dpar{\,(\mdot\: P )}{\mdot} 
-\rho\:  P(\mdot, t) + \rho \,\int_{0}^{\tau\, \mdot} d\,m \: \varphi(m) \; P\left(\mdot - m/\tau, t\right),
\ee
where
\be
\label{e-apar}
\rho = \lambda\,\tau
\ee
\es
is the accretion parameter.
The solution of this equation is unique once we impose the 
normalisation
\be
\label{e-normal}
\int_{0}^{\infty} d\mdot \, P(\mdot, t) = 1
\ee
and a suitable the initial condition.

A dimensional analysis (e.g. \citealp{Barenblatt:1996a}) 
helps to write Eq.~\eqref{e-gfpe} in non-dimensional form.  
Suppose that $\varphi$  depends on the characteristic mass $m_{0}$ as 
the only  dimensional parameter. 
The only dimensional parameters in the distribution $P$ are then
$\mdot$, $t$, $\tau$ and $m_{0}$. 
Note that the clumps' arrival  rate $\lambda$
appears in the equation  only  in the dimensionless combination 
$\rho = \lambda\, \tau$, and therefore it does not appear as a separate
variable. The only way to combine the governing dimensional variables 
into dimensionless variables (from which $P$ must depend) are
\be
x = \mdot\,\tau/m_{0} \qquad  s = t/\tau.
\ee
so 
\be
P = P\left(\frac{\mdot\,\tau}{m_{0}}, t/\tau\right)
\ee
(we do not indicate the dependence on possible other dimensionless parameters).
Plugging this expression into Eq.~\eqref{e-gfpe} we retrieve
\be
\label{e-gfpe-nd}
\dpar{\tilde{P}}{s} = x\,\dpar{\tilde{P}}{x} 
+ (1-\rho) \:  \tilde{P} + 
\rho \,\int_{0}^{x} d\,y \: \tilde{\varphi}(x-y) \; 
\tilde{P}\left(y, s\right),
\ee
where  $\tilde{P} = P\,m_{0}/\tau$
and $\tilde{\varphi} = \varphi\, m_{0}$ are the scaled distributions $P$ and
$\varphi$.
In the following we shall refer to the dimensionless Eq.~\eqref{e-gfpe-nd},
omitting the tildes on the scaled quantities.
  
\bigskip 

The problem \eqref{e-bvp} depends on some initial condition. 
After a long time $t\gg \tau$
the compact star has gathered a large sample of  clumps, 
and the probability $P$ is expected to relax on an equilibrium, 
time independent configuration. 
In the remainder of this paper we will shall assume that that
such an equilibrium has been achieved, and we will work with the
stationary distribution, solution of
\bs
\label{e-station}
\be
\label{e-sgfpe}
x\, \der{P}{x} =
(\rho-1)\, P - \rho \,\int_{0}^{x} d y  \: \varphi(x-y) \; P\left(y\right),
\ee
This is  a Volterra integro-differential equation of the second kind, whose
solution is unique once we impose the  normalisation 
\be
\label{e-norm}
\int_{0}^{\infty} d x \,P(x) = 1.
\ee
\es
The solution to the problem~\eqref{e-station} is particularly
convenient via a  Fourier (or Laplace) transform. This not only allows
a numerical solution, but also yields  some exact results
on the moments of the distribution.

\subsection{Solutions to the generalised Fokker--Planck Equation}
\label{s-solutions}

In this section we derive a solution to the generalised Fokker--Planck
equation.  Although an exact solution does not exists for
a general form of $\varphi$, the Fourier transform of $P$ is
analytical. This allows to compute all the moments
of $P$ as function of the moments of $\varphi$ (when they exist),
and it is makes  possible to compute numerically $P(x)$ 
by evaluating a Fourier integral.

The logarithm of the Fourier transform of Eq.~\eqref{e-sgfpe} reads
\be
\label{e-ft}
\ln  P_{k} = -\rho\, \int_{0}^{k} \frac{d k'}{k'}\:(1- \varphi_{k'}),  
\ee
where $\varphi_{k}$ is  the Fourier transform of  $\varphi$, and where
we have introduced the normalisation~\eqref{e-norm} as $\ln P_{k=0} = 0$.
Plugging $\varphi_{k}$ into the last equation, 
\begin{align}
\label{e-st}
&\ln P_{k} = 
\rho\, \int_{0}^{\infty} d y \: \varphi(y) \:  {\rm Ein}(i\, k\,y),
\intertext{being}
\label{e-ein}
&{\rm Ein}(\zeta) = \int_{0}^{\zeta} d s \: \frac{1 - {\rm e}^{-s}}{s}
\end{align}
the entire exponential integral (see e.g. \citealp{Abramowitz:1972a}). 
The inverse Fourier transform 
\be
\label{e-mprob}
P(x) = \int_{\mathbb R} \frac{d\, k}{2\, \pi} 
\: \exp\left[ - i\, k \, x + \ln P_{k} \right],
\ee
is rather  involved, and cannot be further simplified  for 
a general  $\varphi$. Yet, it is possible to derive some exact results.
The function $\ln P_{k}$ is the 
cumulant distribution function  (CDF) of $P(x)$.
We Taylor-expand the entire exponential integral in 
Eq.~\eqref{e-st}  and compare the resulting expression  with the 
definition of CDF 
\be
\ln P_{k} \equiv \sum_{r=1}^{\infty} \; \frac{(i\,k)^{r} \langle\langle x_{r}\rangle\rangle}{r!}
\ee
to obtain the cumulants
\be
\langle\langle x_{r}\rangle\rangle = 
\rho \; \mean{x^{r}} /r,
\ee
or, in dimensional units, 
\be
\langle\langle\mdot_{r}\rangle\rangle = \frac{\rho}{r} \;
\frac{\mean{m^{r}}}{\tau^{r}}.
\ee
From the cumulants it is easy to retrieve the first moments of
the distribution $P$: in dimensional variables 
\bs
\label{e-moments}
\begin{align}
\mean{\mdot}   &= \lambda \; \mean{m} 
\\
\sigma_{\mdot}^{2} &= \frac{\lambda}{2\, \tau} \: \mean{m^{2}} 
\\
\gamma_{1} &= \frac{2\sqrt{2}}{3\, \rho^{1/2} } \; 
\frac{\mean{m^{3}}}{\mean{m^{2}}^{3/2}}
\\
\gamma_{2} &= \frac{1}{\rho} \; \frac{\mean{m^{4}}}{\mean{m^{2}}^{2}},
\end{align}
\es
where $\mean{\mdot}$, $\sigma_{\mdot}^{2}$, $\gamma_{1}$ and $\gamma_{2}$
are respectively mean, variance, skewness and excess kurtosis of $P$
(see e.g. Sec.~14.1 of \citealp{Press:2007a} for the definitions).

Mean and variance coincide with the values worked out 
in Sec.~\ref{s-stacc} from our simple analysis of Eq.~\eqref{e-sde-1}.
The higher order moments provide further qualitative details on the
shape of the accretion distribution function $P$.
The skewness $\gamma_{1}$ is always positive, meaning that 
$P$ has a long tail for large $x$.
The kurtosis $\gamma_{2}$ is always positive as well, i.e.
$P(x)$ is  leptokurtic: it has a sharp peak and a heavy tail 
for large $x$. 
If the terms $\mean{m^{n}}$ are finite, 
the standardized moment of order $n$ of $P$ scales as
$\rho^{1-n/2}$. 

The computation of the Fourier integral~\eqref{e-mprob} cannot be done analytically for any $\varphi$, but it can be tackled  numerically.  
This will be done later for some choices of
the distribution $\varphi$.

If $\varphi(m)$ is bounded for $m\to 0$, the  asymptotic behaviour of $P$  
for $x\ll1$ is
independent of $\varphi$. For small $x$ the integral in 
Eq.~\eqref{e-sgfpe} is negligible, and  
\be
\label{e-xsmall}
P(x) \sim x^{\rho-1} \qquad \text{for} \qquad x\ll 1.
\ee  
For small values of $x$, $P$ is suppressed if $\rho\gg1$,
and diverges for $\rho\leq 1$. 
This behaviour has a clear physical explanation.
For simplicity, we assume that there is no accretion barrier,
so $P$ may be read as a luminosity distribution.
Consider the accretion of two small clumps, the second
one hitting the system a time $\lambda^{-1}$ after the first one.
The inequality $\rho<1$ means that the relaxation time $\tau$ is 
shorter than $\lambda^{-1}$: the compact object has time to process
the first  clump before the  arrival of the second one, 
and their accretion triggers {\em two} successive flares. 
The accretion of each clump is ``recorded'' in the low luminosity tail, 
and therefore  the value of $P$ may be high. 
On the other hand, if $\rho\gg1$, 
the  compact object cannot process the first clump before the arrival of the second one. The accretion of two closely separated clumps cannot trigger {\em two} different low luminosity flares. Such accretion episodes 
are ``recorded'' as the accretion of a singe larger 
clump, and the low luminosity tail of $P$ is depressed by this effect
that for convenience will be referred to as ``$\rho$ suppression''.

\subsection{A Dirac delta  mass clumps' distribution}
\label{s-delta}

\begin{figure}[ht]
\centering
\includegraphics[width=110mm]{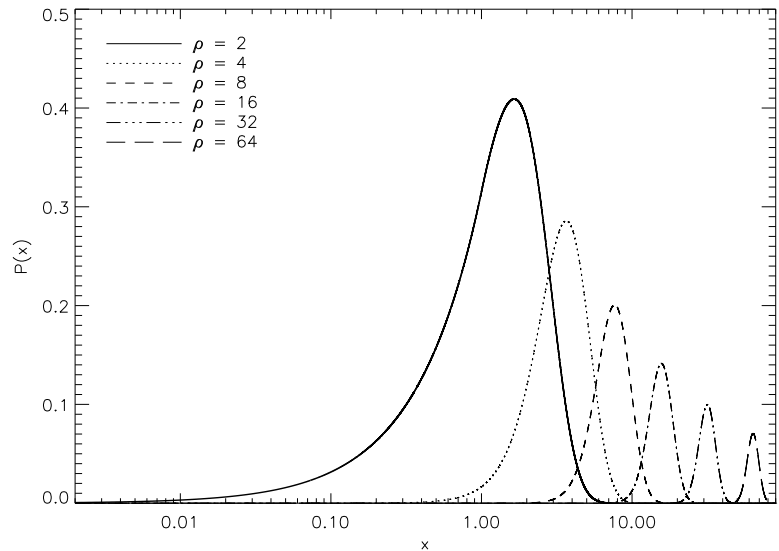}
\caption{\label{f-delta}
The accretion PDF for the delta clumps' distribution~\eqref{e-delta}
for several values of the accretion parameter $\rho$.
}
\end{figure}

Before applying the present model to a concrete astrophysical problem, 
it is useful to sketch the properties of $P$ in the relatively
simple case in  which all the clumps have the same mass.
The appropriate mass distribution is then
(in dimensionless units)
\be
\label{e-delta}
\varphi(x) = \delta(x - 1),
\ee
where $\delta$ is the Dirac delta function.
The generalised Fokker--Planck equation~\eqref{e-sgfpe} 
reduces to the  delay differential equation
\be
\label{e-delay}
x\, \der{P}{x} = (\rho-1)\, P - \vartheta (x-1) \; P(x-1),
\ee
where $\vartheta(u)$ is the Heaviside step function, 
$\vartheta=0$ for $u<0$ and
$\vartheta=1$ for $u>0$.
The solution of Eq.~\eqref{e-delay} can be easily tackled numerically.
Some plots of $P(x)$ are shown in Fig.~\ref{f-delta}
for several values of the accretion parameter $\rho$. 

The mean of $P$  is 
$\mean{x} = \rho$ and the standard deviation is $\sigma = \sqrt{\rho/2}$.
On account of the lack of massive clumps in the 
distribution~\eqref{e-delta}  the distribution $P$ does not display a significant tail. If $\rho>1$, $P$ is also  suppressed
for $x\lesssim 1$.
These effects make the distribution $P$ quite narrow around its mean.
The limit distribution solution of~\eqref{e-delay} will be useful
later as a benchmark for more complex cases.

\subsection{A log-normal clumps' mass distribution}
\label{s-lognormal}

We now explore in some detail the solution of  the GFPE~\eqref{e-gfpe} with
a more general distribution of the clumps' masses. 
To our purposes, it is particularly convenient to adopt 
the log-normal distribution
\be
\label{e-lnorm}
\varphi(x) = \frac{1}{x\, \sigma\, \sqrt{2\,\pi}} \; 
{\exp}\left[{-\frac{\ln^{2}x}{2\, \sigma^{2}} }\right],
\ee
defined for any  $x>0$.
This choice has several advantages. First, 
the log-normal  is extremely flexible: by tuning the shape parameter 
$\sigma$, it 
may mimic distributions as different as a Dirac delta (for $\sigma\ll1$), 
or a power law (for  $\sigma \gtrsim 1$). The distribution
\eqref{e-lnorm} may be written (up to a constant factor)
$\varphi\sim x^{-\zeta}$, where $\zeta = 1 + \ln x / 2\,\sigma^{2}$.
The exponent $\zeta$ is a slowly varying function of the range
$x$, and for $x\ll {\rm e}^{{\sigma^{2}/2}}$ the log-normal 
$\varphi$ closely resembles a power law with index $\zeta\gtrsim 1$.
All the  moments of a log-normal are defined, being 
\be
\label{e-lnmom}
\mean{x^{n}} = \exp{\left(n^{2}\, \sigma^{2} / 2\right)}.
\ee
Eqns.~\eqref{e-moments} and~\eqref{e-lnmom} give the
moments of the  distribution $P$
\bs
\label{e-lognmoments}
\begin{align}
\mean{\mdot}   &= \lambda \,  m_{0} \; {\rm e}^{\sigma^{2}/2}
\\
\sigma_{\mdot}^{2} &= \frac{\lambda}{2\,\tau} \; m_{0}^{2} \; 
{\rm e}^{2\, \sigma^{2}}
\\
\gamma_{1} &= \frac{2\sqrt{2}}{3\, \rho^{1/2}} \; {\rm e}^{3\,\sigma^{2}/2}
\\
\gamma_{2} &= {\rm e}^{4\, \sigma^{2}}/\rho.
\end{align}
\es
It is clear that the moments of $P$ are defined for any order $n$, 
but their  magnitude rapidly increases with $n$ if $\sigma >1$,
making $P$ heavy tailed.

The Fourier integral~\eqref{e-mprob} cannot be computed exactly 
if $\varphi$ is a log-normal, but it must be evaluated numerically.
The numerical calculation of Fourier integrals poses a well known
 problem, widely discussed in the mathematical literature.
A very convenient numerical method is a variant of
the double exponential quadrature
(see e.g. \citealp{Mori:2001a})  suggested by~\citet{Ooura:1999a}.
%
\begin{figure} 
\centering
\includegraphics[width=90mm, angle=90]{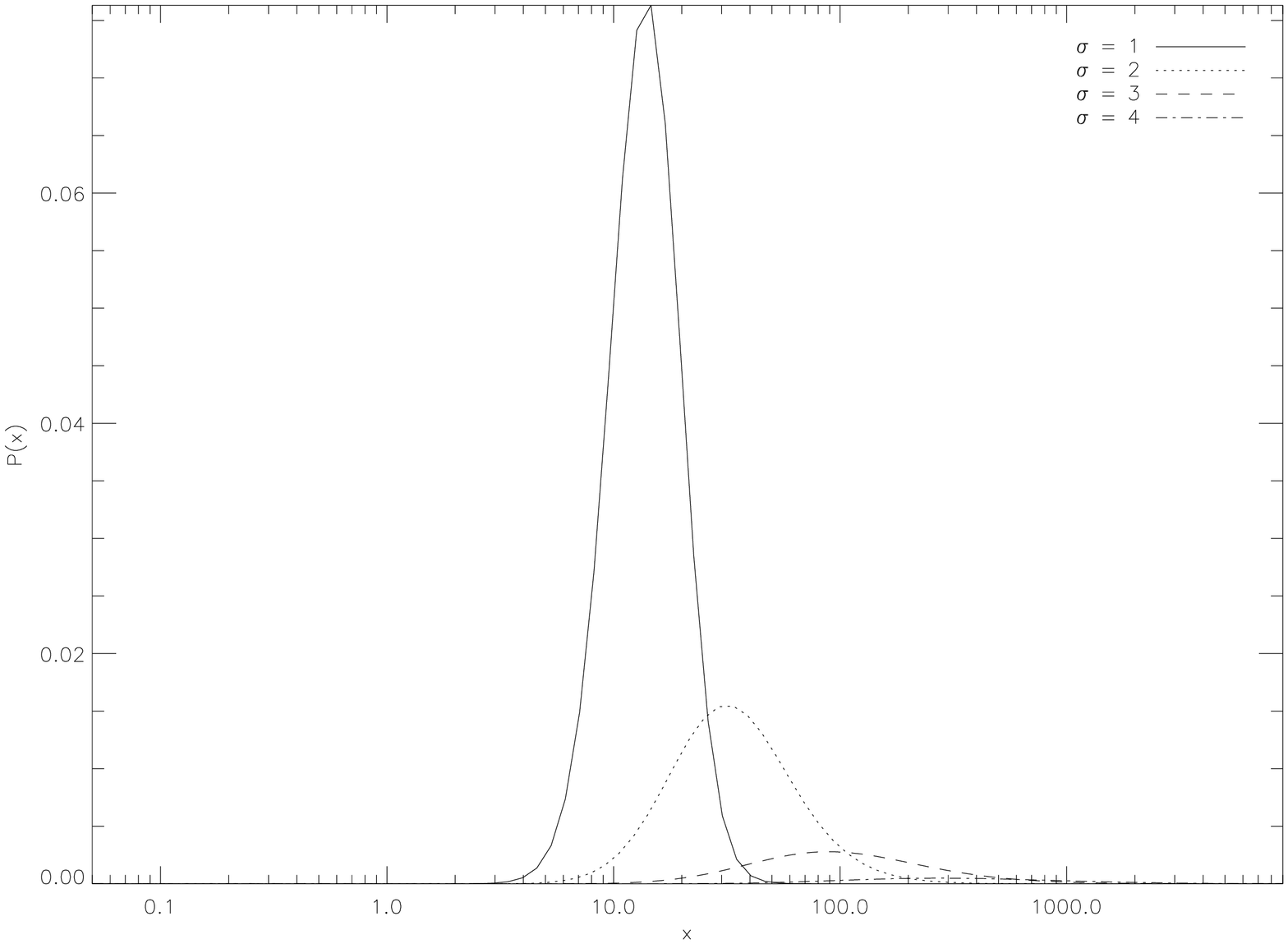}
\includegraphics[width=90mm, angle=90]{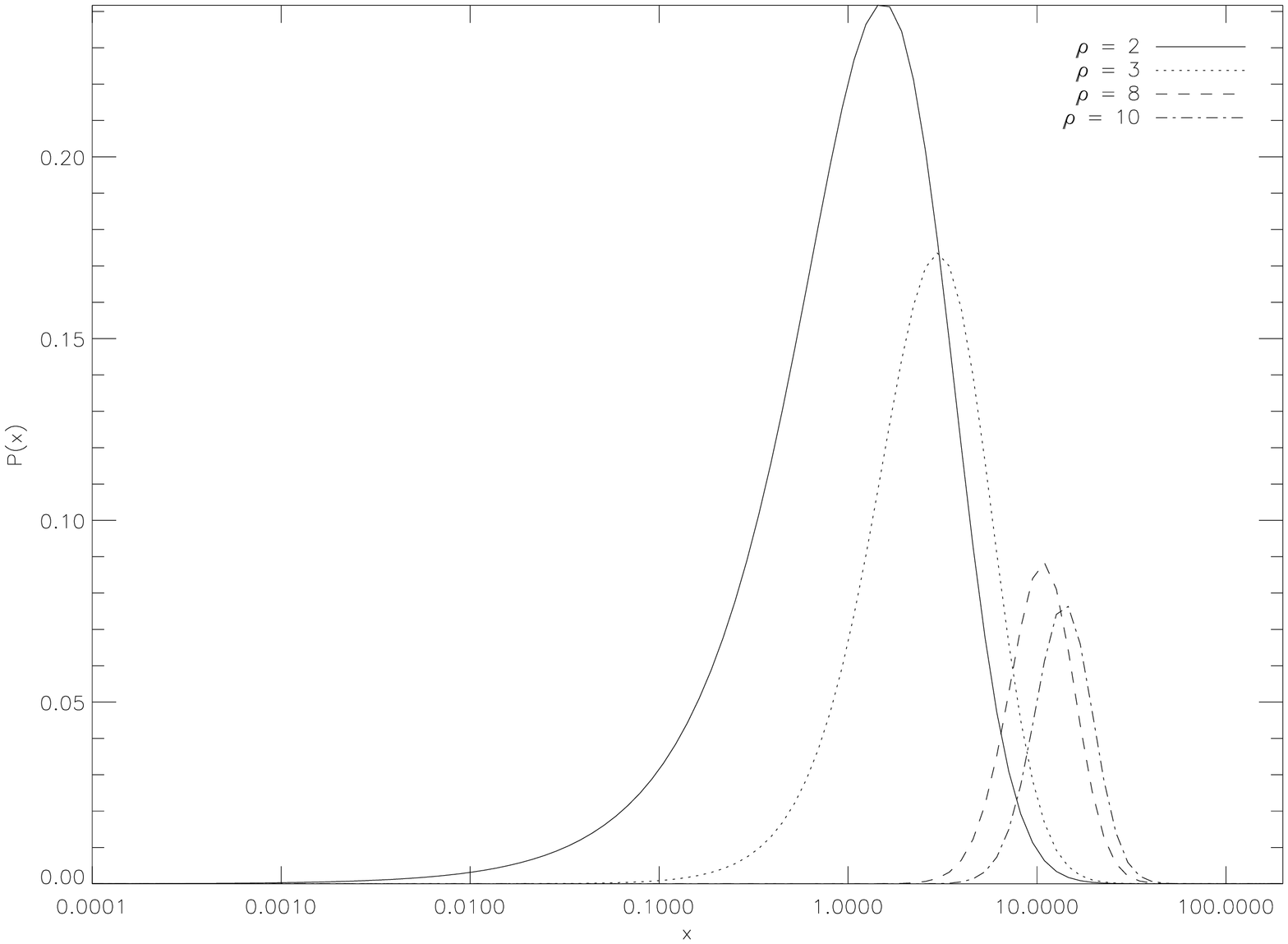}
\caption{\label{f-plots-ln}
Two samples of theoretical curves computed from the log-normal
distribution~\eqref{e-lnorm}. 
In the top panel $\rho$ is fixed  at $\rho=10$
and the curves refer at different values of $\sigma$.
In the  bottom panel $\sigma$ has been fixed at $\sigma=1$
and the curves refer to different values of $\rho$.
}
\end{figure}

%
In Fig.~\ref{f-plots-ln}   we show some theoretical curves computed
from a log-normal $\varphi$ for several values of $\sigma$ and
$\rho$.
If $\rho$ is fixed (see the top panel of Fig.~\ref{f-plots-ln}),
as $\sigma$ increases, also the distribution becomes 
increasingly broad, stretching its tail to high values of 
$x$. 
The peak of $P$ moves to the right with increasing $\sigma$, and
its height lowers to preserve  the normalisation. 
For moderately large values of $\sigma$, the interval over which 
$P$ significantly differs from zero  spans several orders of
magnitude.

For a fixed $\sigma$, on the other hand, 
(bottom panel of Fig~\ref{f-plots-ln}) for increasing values of $\rho$
the distribution $P$ is suppressed for low $x$. This general
property of $P$ has been already discussed at the end of 
Sec.~\ref{s-solutions}.


\section{Application to Supergiant Fast X--ray Transients}
\label{s-sfxt}

We apply the formalism developed in this paper to the analysis of the flares' distribution observed in the Supergiant Fast X--ray Transients (SFXTs). 
We first outline the main properties of this class of objects, and
then will analyse them in the framework of stochastic accretion.

\subsection{Presentation of SFXTs}

SFXTs   are a class of rare
transient X--ray sources  (about ten members are known up to date),
associated to OB supergiant stars  (see \citealp{Sidoli:2011b} for a recent review), making them a sub-class of the HMXBs.

The short and bright X--ray emission from several members of this class 
was discovered by the {\sl INTEGRAL} satellite during monitoring observations of the Galactic plane at 
hard X--ray energies (\citealp{Sguera:2006}, \citealp{Negueruela:2005a}).
About a half of them are X--ray pulsars and the mass flow from the massive donor 
to the compact star (usually assumed to be a neutron star in all the members of the class) 
occurs via a stellar wind.

SFXTs spend most of their time in a relatively low  level 
brightness state (below $L_{X}\sim 10^{34}\ergs$; \citealp{Sidoli:2008a}), 
reaching $L_{X}\sim 10^{32}\ergs$ in quiescence, but show
an occasional ``outbursting'' activity  when
the mean X--ray luminosity is much higher than usual for a few days,
and is punctuated by a sequence of short (from a few minutes to a few hours long) flares 
peaking at X--ray luminosities of   $10^{36}-10^{37}\ergs$ 
(e.g. \citealp{Romano:2007, Rampy:2009}).

Their X--ray transient activity with such a wide dynamic range (from 2 to 5 orders of magnitude) 
is puzzling, since the components of these binary systems seem very similar
to persistently accreting classical X--ray pulsars (e.g. Vela X--1).  
Two  kind of explanations have been proposed, although none of them
is able to completely explain the whole phenomenology of {\em all} the members of the class, to date.

In the first type of models, the compact object accretes
mass from an extremely inhomogeneous ``clumpy'' stellar wind \citep{Zand:2005}, which is
believed to be present in massive OB stars
(see e.g. the review of \citealp{Puls:2008a}).
In a few cases a preferential plane for the outflowing clumpy 
wind is suggested (\citealp{Sidoli:2007}, \citealp{Drave:2010}) while usually
a spherically symmetric morphology is assumed \citep{Negueruela:2008}. 
It is still unclear if the flow is able to form a (possibly transient) 
accretion disc (see e.g. \citealp{Ducci:2010a}). 

In the second kind of models, the compact object is a slowly spinning 
magnetic neutron star undergoing transition
between the accretion regime and the onset of a centrifugal or a 
magnetic barrier (\citealp{Grebenev:2007}, \citealp{Bozzo:2008a}).
This scenario involves the interplay of a host of physical processes, 
including several kinds of  hydrodynamical instabilities the details of
which are beyond the simplified sketch outlined in this paper.

\begin{figure}[ht]
\centering
\includegraphics[width=110mm]{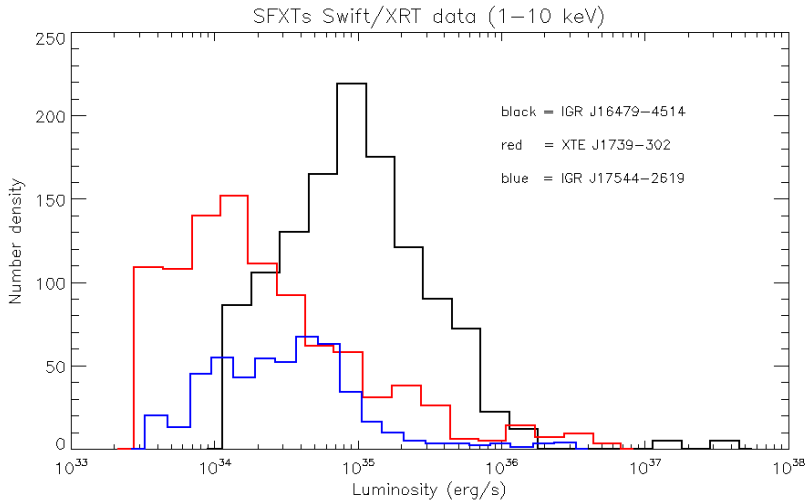}
\caption{\label{f-histo}
Distribution of the X--ray luminosity of three SFXTs  
we  calculated from the
X--ray count rate distribution observed  with 
{\sl Swift}, during  a two years' campaign. Data have been 
taken from \citet{Romano:2010a}.
}
\end{figure}

A sample of SFXTs has been extensively observed by the X--ray satellite 
{\sl Swift} \citep{Romano:2010a}. 
Fig.~\ref{f-histo} shows the  histograms of the flares' luminosity distributions for three SFXTs  that we computed from the data gathered in a two years' campaign with {\sl Swift} \citep{Romano:2010a}. 

These histograms are close relatives to the  
flares' luminosity distribution FLD introduced in Sec.~\ref{s-intro}. 
We derived the X--ray luminosity histograms from the distributions
of count rates originally reported in literature by \citet{Romano:2010a}.
Conversion factors from  {\sl Swift} XRT count rates to X--ray luminosities ($1-10~\keV$) 
have been calculated assuming a power law
spectrum, with the distances and the 
spectral parameters appropriate for each individual SFXT, as reported in 
\citet{Sidoli:2008a}.
%


\subsection{The model}
\label{s-model}

In this section we interpret the observed luminosity distributions 
plotted in Fig.~\ref{f-histo} in the light of the stochastic 
set up in the previous sections. 
Some of the questions we address are the following:
\begin{itemize}
\item
The  SFXTs' luminosity distribution drops quite abruptly 
below $L_{X}\simeq 10^{33}-10^{34}\ergs$. 
Provided that this is not a selection effect in our observations, what is the origin of this feature?
\item
What properties of the accretion stream and accretion process
can we infer by comparing our model with the observations?  
How do the properties 
of the stellar wind we deduce compare with the existing 
literature?
\item
Can we explain the difference between the wide dynamic range 
of luminosities observed
in SFXTs and the much narrower one seen in persistent sources, like e.g.
Vela X-1? 
\end{itemize}

Our comparison of the model with the observed flares' luminosity
distributions (FLDs) is not intended to be a 
``best fitting'' procedure, for several reasons.

First of all, our model~\eqref{e-window} for the response of the 
accretor may be a bit oversimplified.   The theoretical 
luminosity distribution $P$ computed from it, therefore, 
is not expected to provide an accurate model of the SFXT flares' 
luminosity distribution observed in the real world
\footnote{Maybe the response~\eqref{e-window} is not
always thus bad. The SFXT $IGR~J16479-4514$, for instance,
is known to exhibit flares that after a fast rise decay
exponentially \citep{Sguera:2005a, Ducci:2010a}.
\citet{Sguera:2005a} could fit the light curve of a flare with
sufficient statistics with an exponential law 
with an e-folding time $\tau=15\pm9~\rm min$.
}.
Second, only the FLDs of three sources are
available for our analysis, probably too few to draw
definite conclusions. Third, not all the flares may have been 
spotted during the observation campaign, and  the FLDs 
we work with might  not describe a fair sample of all the flares. 
Finally, the calculation of a model for given values of $\rho$ and $\sigma$
is quite time consuming, and a quantitative determination of
those parameters via a best fitting procedure is not viable 
at the moment. 

When we compare an observed FLD with its theoretical model, we aim 
to get a hint at the properties of the accretion process 
(described  by the parameter $\rho$) and the properties of the family 
of clumps (described by the log-normal distribution~\eqref{e-lnorm}).
A more physically motivated response is clearly necessary in order to make 
more  quantitatively definite  statements.

Despite its all too obvious limitations, we believe that our model 
has its own value. Its qualitative predictions are expected to hold even for
a more sophisticated model; in addition it points out the
fundamental properties of stochastic accretion  that 
can be easily shaded by rather opaque  mathematical complications
that we leave to a forthcoming paper.
That said, we are now ready to compare out theoretical predictions
with the observed flares' distributions in the three SFXTs 
IGR~J$16479-4514$, IGR~J$17544-2619$ and XTE~J$1739-302$.

 
We aim at determining  the best fitting  parameters of our model 
from the flares' distributions  of a given SFXT. 
In general, the free parameters of the model are the accretion parameter 
$\rho$, the log-normal shape factor $\sigma$, the 
ratio $m_{0}/\tau$ and the cut-off mass accretion
rate $\mdot_{\omega}$  due to the action of an accretion
barrier (see Eq~\eqref{e-lum-propeller}). In order to keep things as simple as possible, in the following analysis we assume that no accretion barrier is present 
(i.e. $\mdot_{\omega}=0$),  
and the the model is completely defined by the three 
parameters  $\rho$, $\sigma$ and $m_{0}/\tau$.

What is the best way to determine these parameters from the data of
an observation campaign of a given SFXT? As we shall see, our choice is
rather limited. 

One might think, for instance, to determine the parameters by comparing
the mean, variance, skewness and excess kurtosis computed from our model
(see Eq.~\eqref{e-moments})  with the observed mean, variance, skewness and excess kurtosis, provided that we are able to convert the mass accretion
rate $\mdot$ to a X--ray luminosity. If there is no accretion barrier,
the modelled luminosity distribution is  simply $p(L) = P(x) / L_{\ast}$, 
where $P(x)$ is the dimensionless distribution computed from the 
generalized Fokker--Planck equation, and
\be
\label{e-lstar}
L_{\ast} = \frac{G\, \mns\, m_{0}}{\rns\, \tau},
\ee
where $\mns\sim 1.4\msun$ and $\rns\sim 10~\rm km$ are the typical mass and
radius of a neutron star.
A direct comparison of the moments of the theoretical and the computed
distributions, however, is not viable. The use of
high order momenta such as the skewness or the kurtosis of a heavy tailed
distribution (like our $P(x)$)  is not recommended for any statistical
analysis (e.g. \citealp{Press:2007a}). The problem is that these moments are not
robust, i.e. they are too sensitive to the outliers. Any small change in 
the number of the observed high luminosity flares would alter the
values of these moments considerably, making them of little use to any
statistical purpose. 
 
Since we cannot use the moments, we are forced to determine the 
parameters by comparing the full shapes of the observed and the theoretical
distributions.  The data from \citet{Romano:2010a} available to our
analysis are  already binned in luminosity intervals: we known
the fraction $P_{i}^{(\rm obs)}$ of the flares observed in the
luminosity interval $[L_{i}, L_{i+1}$],  but we don't have 
access to the luminosity of each single flare. 
This prevents us from using  any variant of the Kolmogorov--Smirnov test
to compare the theoretical and the observed flares' distributions.

For these reasons, we must resort to a least squares method to determine the
parameters $\rho$, $\sigma$ and $m_{0}/\tau$.  Suppose for the moment
that $\rho$ and $\sigma$ are known, and the only parameter to be
determined is $m_{0}/\tau$. For a (tentative) value of $m_{0}/\tau$ we 
compute the scale luminosity~\eqref{e-lstar} and the expected fraction 
$P_{i}^{({\rm cal})}$ of flares within the luminosity interval 
$[L_{i}, L_{i+1}$]. 
We compute the chi-squared
\be
\mathcal{X}^{2} = \sum_{i} 
\frac{\left[P_{i}^{(\rm obs)} - P_{i}^{(\rm cal)}\right]^{2}}{\sigma_{i}^{2}},
\ee
where we assume the errors on the observed fractions
$\sigma_{i}^{2} \sim P_{i}^{(\rm cal)}$. Finally,
we find the value of $m_{0}/\tau$ that minimises $\mathcal X^{2}$. 
A minimisation procedure to determine all the parameters $\rho$,
$\sigma$ and $m_{0}/\tau$ would be very time consuming, and therefore is
not viable at the moment. For this reason we have  built a grid of
values of $\rho$ and $\sigma$, and for each couple ($\rho$, $\sigma$)
we have computed the correspondent $P(x)$, and the value of
$m_{0}/\tau$ that minimises the $\mathcal X^{2}$ for that $P$.
Finally,  we chose as our best estimates for $\rho$, $\sigma$
and $m_{0}/\tau$ the values that return the smallest  
$\mathcal{X}^{2}$. On account of the (expected) poor quality of
the fits, we do not report the values of the 
$\mathcal{X}^{2}$ statistics,
but we only discuss here below our results.

%

\begin{figure} 
\centering
\includegraphics[width=65mm,angle=90]{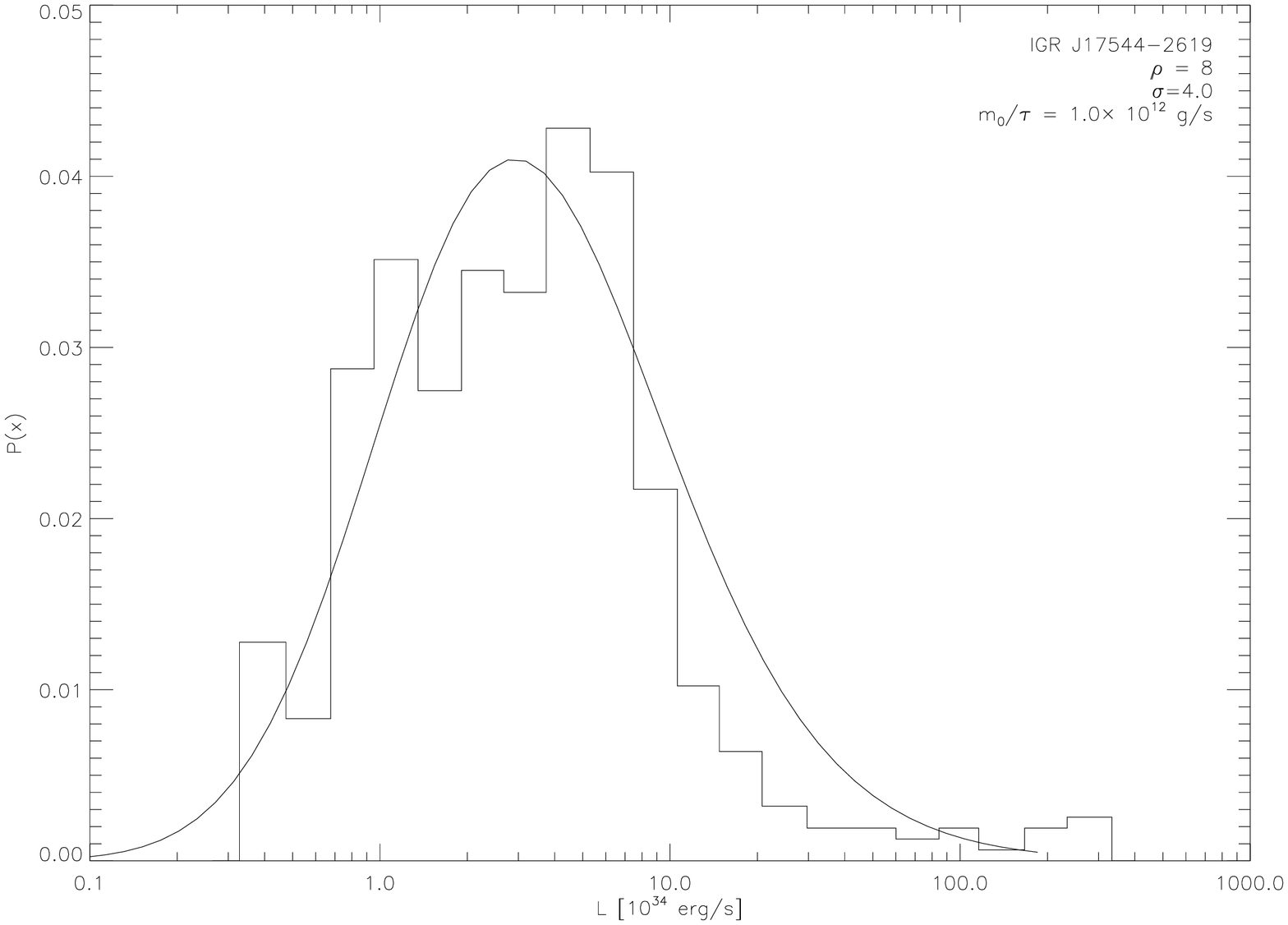}
\includegraphics[width=65mm,angle=90]{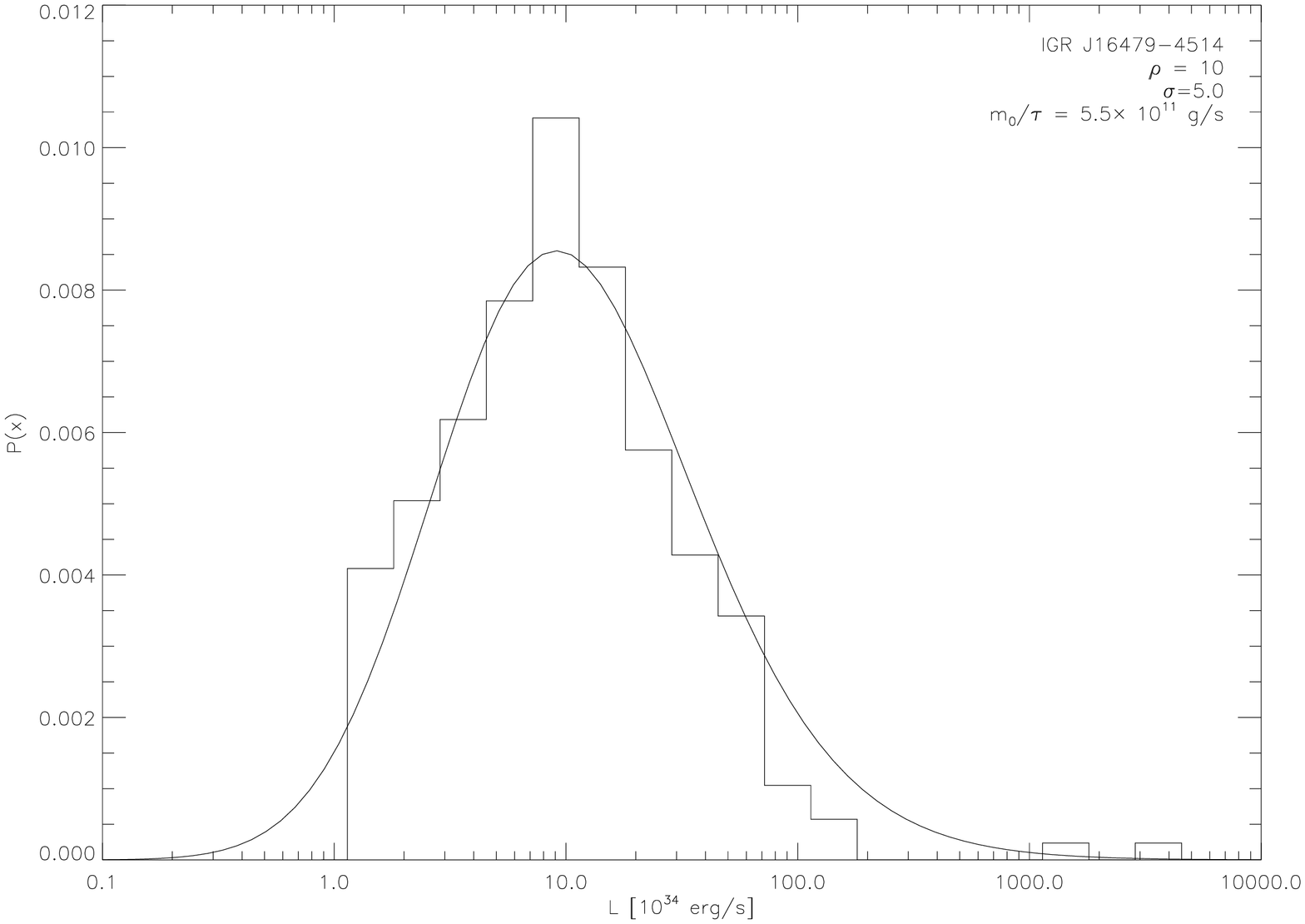}
\includegraphics[width=65mm,angle=90]{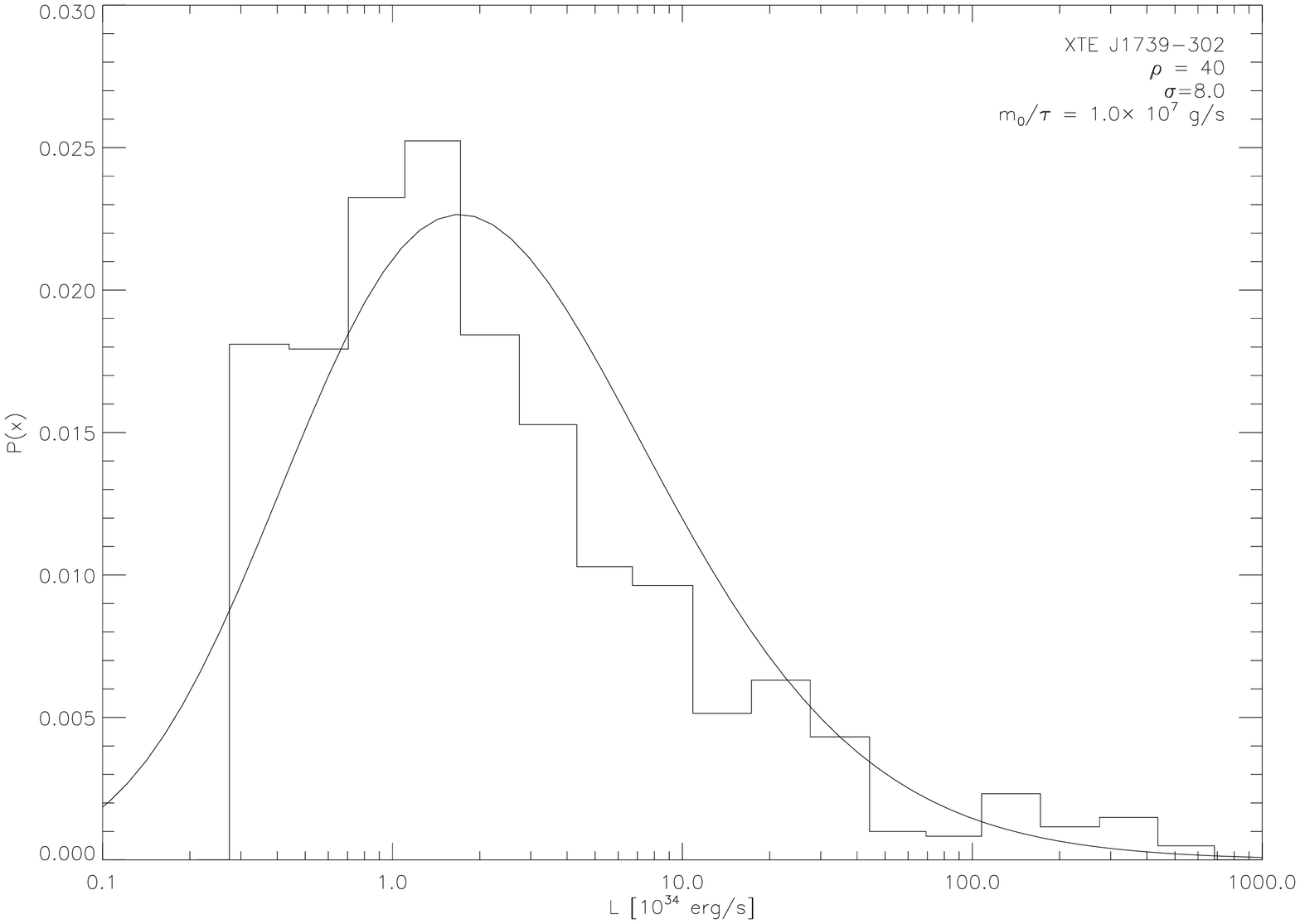}
\caption{\label{f-fits}
The observed FLDs of the SFXTs 
$IGR~J17544-2619$ (top panel),
$IGR~J16479-4514$ (middle panel) and
$XTE~J1739-302$ (bottom panel)
superposed to their 
``best fitting'' theoretical curves.
The closest agreement between the observed and the computed FLD
are found for 
$\rho=8$, $\sigma=4$ and $m_{0}/\tau = 1.0\times 10^{12}\g\s^{-1}$.
(for $IGR~J17544-2619$),
$\rho=10$, $\sigma=5$ and $m_{0}/\tau = 5.5\times 10^{11}\g\s^{-1}$
(for $IGR~J16479-4514$) and
$\rho=40$, $\sigma=8$ and $m_{0}/\tau = 1.0\times 10^{7}\g\s^{-1}$
(for $XTE~J1739-302$).
}
\end{figure}

Fig.~\ref{f-fits}  shows the flares' distributions 
for our  three SFXTs superposed to their ``best fitting'' models.
On account of the  asymptotic behaviour~\eqref{e-xsmall} 
of the theoretical luminosity distribution, a glance to
Fig.~\ref{f-histo} hints that
in all cases it must be $\rho > 1$ to match the observed FLDs.
In addition, the observed  high dynamical ranges of the 
flares' luminosities require a heavy tailed clumps' mass
distribution, i.e. a relatively high log-normal shape parameter $\sigma$. 
%

$IGR~J17544-2619$ is the source with the lowest dynamic range,
since it displayed flares ranging in the luminosity interval
between $3.9\times 10^{33}\ergs$ and $2.7\times 10^{36}\ergs$.
There is no apparent low luminosity cut--off 
(which might be interpreted as the effect of an accretion barrier).
The FLD of this source may be reproduced quite nicely by a model 
with $\rho=8$, $\sigma=4$ and
$m_{0}/\tau=1.0\times 10^{12} \g\s^{-1}$.

The source $IGR~J16479-4514$ has the highest  dynamic range
of all the considered sources,
with flares' luminosities ranging from $1.4\times 10^{34}\ergs$
to $3.5\times 10^{37}\ergs$.
In this case the best value of the log-normal shape is
$\sigma=5$ (slightly larger than the value found for $IGR~J17544-2619$).
The best estimate of the accretion parameter is
$\rho = 10$, and even in this case  it is not probably necessary to
invoke an accretion barrier to explain the left tail of the observed FLD.
The last parameter is $m_{0}/\tau=5.5\times 10^{11}\g\s^{-1}$.

The flares of $XTE~J1739-302$ show luminosities between $3.3\times 10^{33}\ergs$
and  $5.3\times 10^{36}\ergs$. 
Its steep ``wall'' that cuts off 
$S(L)$ below $L=2.1\times 10^{33}\ergs$ is difficult to reproduce. 
Following our discussion of the ``$\rho$ suppression'' at the 
end of Sec.~\ref{s-solutions}, one might expect that 
a very high value of $\rho$ is necessary to cut off abruptly 
the low tail of the luminosity distribution.
Indeed, our best-fitting model for this source gives $\rho=40$,  $\sigma=8$ 
and $m_{0}/\tau = 1.0\times 10^{7}\g\s^{-1}$.
We note however that the space of parameters characterised by high 
values of $\rho$ and $\sigma$ is quite difficult to explore with our means.
For this reason, and also on account of the relatively poor quality
of the ``best fit'' shown at the bottom panel of Fig.~\ref{f-fits}, 
we cannot be sure that the $\rho$ suppression 
is winning over the  ``propeller barrier'' process discussed in  
Sec.~\ref{s-dsde}.

To resume, we may reproduce at least the gross properties of 
the sources' FLDs, with  moderate values of the accretion
parameter $\rho$ and  a relatively high
shape parameters $\sigma$ of the clumps' log-normal. 
At least two of our sources do not require any accretion barrier
to explain their low luminosity tail, that may 
be justified by the ``$\rho$ suppression''. The source $XTE~J1739-302$
is more problematic, and here a propeller barrier is not excluded.

\bigskip 

Our model has three degrees of freedom: 
$m_{0}/\tau$, $\sigma$ and the accretion parameter 
$\rho\equiv \lambda\, \tau$.
The  mass arrival rate $\lambda$ is degenerate since it only
appears in a product with $\tau$ in the accretion
parameter $\rho$, and so it cannot be determined directly from a
comparison of our model with the observations.
We may however get a hint from the analysis of
the flares' light curves and from the geometry of the system.

In their analysis of the light curve of 
$IGR~J16479-4514$  \citet{Sguera:2005a} 
found that the flares lasted from $\sim30~\rm min$ to
$\sim3~\rm h$.
In one case, they were able to analyse 
a single light curve that after a fast rise,
decayed exponentially with an e-folding time
$\tau = 15\pm9~\rm min$. Exponentially decaying light
curves are not uncommon for the flares occurring 
in this source \citep{Ducci:2010a}.
If the decay time $\tau\sim 10^{3}\s$ may be taken
as typical, then for $\rho\sim 10$ we estimate
the time interval between the successive accretion of two clumps
$\lambda^{-1}\sim 10^{2}\s$. If the NS orbits a massive star
of $M\simeq 20\msun$ with a period  $P_{\rm orb}\simeq 3~\rm d$
(typical of $IGR~J16479-4514$), 
the distance covered by the NS between two successive encounters
is of the order of few $10^{9}\cm$, a fraction of the stellar radius.
From the values of $m_{0}/\tau$ and $\tau\sim10^{3}\s$ we find
values of $m_{0}$ between $\sim10^{10}\g$ and $\sim 10^{15}\g$
for our three sources.
On account of the large values of $\sigma$, the mean masses 
are much larger than these
(compare with Eq.~\eqref{e-lnmom}),
and are of the order of $10^{18}-10^{21}\g$. 
These values for the clumpy wind are in broad agreement
with estimates obtained with other means 
(see e.g. \citealp{Ducci:2009a}).

\bigskip
 
We now present a simple argument showing  how 
the parameters  $m_{0}/\tau$, $\sigma$ and $\rho$
may  not be fully independent of each other. 
Consider a giant donor star with mass $M_{\star}$
and radius $R_{\star}$ that blows a clumpy wind. Assuming that
the mass carried by the wind is all in clumps, then
the mean stellar mass outflow rate at the distance $r$ from the star is
\be
\label{e-mwind}
\mdot_{w} = 4\,\pi\, r^{2} \, n_{\rm cl} \, v_{\rm cl} \, \mean{m},
\ee
where $n_{\rm cl}$ is the mean number of clumps per unit volume, and 
$v_{\rm cl}$ is the mean clumps' velocity.
As the neutron star orbits the giant star, it  crosses this 
clumpy wind and encounter clumps with the mean rate
\be
\label{e-lambda}
\lambda = \pi \, R_{G}^{2} \, v_{\rm orb} \, n_{\rm cl},
\ee
where $v_{\rm orb}$ is the orbital velocity of the NS and $R_{G}$
is the capture radius defined by Eq.~\eqref{e-rcapt}. 
Eliminating $n_{\rm cl}$ from Eqns.~\eqref{e-mwind} and~\eqref{e-lambda} 
and postulating that the population of the masses is log-normally distributed
with shape parameter $\sigma$ and median mass $m_{0}$
we find
\be
\label{e-mlambda}
\lambda = \left(\frac{R_{G}}{2\, r}\right)^{2} \;
\frac{\mdot_{w}}{m_{0}} \; 
\frac{v_{\rm orb}}{v_{\rm cl}}
\; {\rm e}^{-\sigma^{2}/2} 
\ee
The rate $\lambda$ is a strongly decreasing  function of the clumps' mass distribution shape $\sigma$ for the following simple reason.
If $\sigma$ is large, the log-normal distribution is skewed
to the right, and  most of the mass blown by the wind  is carried by large, heavy clumps. A given $\mdot_{w}$ is carried by  relatively few clumps, and their  encounters with  the orbiting  neutron star are rare, i.e.
$\lambda$ is small.  
Multiplying  the last equation by $\tau$, we find
\be
\label{e-mrho}
\rho = \left(\frac{R_{G}}{2\, r}\right)^{2} \;
\frac{\mdot_{w}}{m_{0}/\tau} \; 
\frac{v_{\rm orb}}{v_{\rm cl}}
\; {\rm e}^{-\sigma^{2}/2}. 
\ee
This shows that in our model the parameters $\sigma$, $\rho$
and $m_{0}/\tau$ are not independent of each other:
$\rho$ is proportional to  $(m_{0}/\tau)^{-1}$
and decreases exponentially with $\sigma^{2}/2$.
It is difficult to use this equation to make quantitative predictions
about the magnitude of $\rho$ as a function of the other 
parameters. Indeed,  when made explicit,
its right hand side is found to  be very sensitive to
the values of  rather uncertain quantities 
(e.g. the velocity of the stellar wind), and even a small change in their
value may vary $\rho$ by orders of magnitude.  
We may plug the   values of $\lambda$ and  $\rho$ into 
Eqns.~\eqref{e-lognmoments} to find 
\be
\label{e-dispersion}
\frac{\sigma_L}{\mean{L}} = \left(2 \:\frac{m_{0}/\tau}{\mdot_{w}}\; \frac{v_{\rm cl}}{v_{\rm orb}}\right)^{1/2} \;
\frac{r}{R_{G}} \; {\rm e}^{3\sigma^{2}/4},
\ee
where $\mean{L}$ and $\sigma_{L}$ are respectively the  mean
accretion luminosity and the standard deviation of $L$, if $L\propto \mdot$.
%
The exponential factor at the right hand side of
the Eq.~\eqref{e-dispersion} suggests an explanation for the difference
between the ordinary, persistent X--ray sources occurring in 
the high mass binaries, and the SFXTs.
Indeed, according to Eq.~\eqref{e-dispersion}, 
even moderate values of $\sigma$ 
(say, $\sigma\simeq 4-5$) greatly amplify the dispersion around the 
mean $\sigma_{L}/\mean{L}$.
This means that the wide dynamical range of luminosities observed in
SFXTs is but the effect of the accretion of a 
population of clumps with a wide mass spectrum $\varphi(m)$.
On the other hand, if $\sigma\ll1$, i.e. all the accreting clumps
have very similar masses, there is no such strong 
amplification, and $\sigma_{L}/\mean{L}$ may be small.
The FLD generated by the accretion of clumps with
narrowly distributed masses should  resemble
the distribution $P$ discussed in Sec.~\ref{s-delta}.

According to this model, then, the difference between persistent
HMXBs and SFXTs is  entirely due to a substantial difference between
the winds blown by the donor stars hosted in these systems. 
Since  SFXTs are much rarer than the ordinary HMXBs, probably 
the occurrence of  a clumpy stellar wind with a wide range of 
 masses is also a rare event.


\section{Summary and conclusions}
\label{s-summary}

In this paper we have set up a simple stochastic model for the accretion
of a clumpy stream  on a compact object. 
We have modelled the response of the compact object to the
accretion with a simple exponential law~\eqref{e-window} 
characterised by the relaxation time $\tau$.
The accretion stream is described as a train of pulses
hitting the compact object with the mean rate $\lambda$.
The clumps have a mass distribution described by the
function $\varphi(m)$. 
With these ingredients we have written a stochastic differential
equation (SDE) for the mass accretion rate $\mdot(t)$.
We have also derived the generalised Fokker--Planck equation (GFPE)
associated to the SDE. Its solution $P(\mdot, t)$ is the probability
that the mass accretion rate is $\mdot$ at the time $t$. 
We have then restricted us to the analysis of the stationary
solution $P(\mdot)$. In general $P$ cannot be written
in a closed form for any choice of $\varphi$, but it is
nevertheless possible to express its moments as functions of
the moments of $\varphi$ (if they exist).
In addition, the asymptotic behaviour of $P(\mdot)$ for small 
$\mdot$ is $P(\mdot)\propto \mdot^{\rho-1}$, for any 
$\varphi(m)$ limited for small $m$.  
For $\rho\gg1$, then, $P(\mdot)$ is negligible for
$\mdot\lesssim m_{0}/\tau$. This ``$\rho$ suppression'' of $P$ 
occurs when the relaxation time $\tau$ is long
with respect to the clumps' arrival rate, and the compact object cannot
respond promptly enough to the rapid succession of clumps' arrivals 
(Sec.~\ref{s-solutions}).

We have then studied in some detail the function $P$
for two choices of $\varphi$: a delta function and a log-normal
distribution.
In the case of a log-normal $\varphi$, $P$ depends on 
three parameters:  the 
accretion parameter $\rho\equiv\lambda\,\tau$, 
$m_{0}/\tau$ and  $\sigma$, where $m_{0}$ and $\sigma$ are the location
and shape parameters of the log-normal. 

As a case study, we have applied our formalism (with a log-normal 
$\varphi$)  to the physics
of Super Giant Fast X--ray Transients (SFXTs), a peculiar sub-class of
High Mass X--ray binaries (HMXBs).

We have compared our model to the flares' X-ray luminosity distributions of
the  SFXTs  $IGR~J17544-2619$, $IGR~J16479-4514$ and
$XTE~J1739-302$  observed in a two years' campaign with {\sl Swift}.
 
We have found that, however simplified, our model may reproduce the
features of the observed flares' distributions.
The typical parameters giving the best agreement between the 
model and the observations are $\rho\sim 10$ and $\sigma\sim5$,
corresponding to a long tail in the mass distributions.
The parameters $\tau$ and $\lambda$ cannot be determined
separately, but taking the value $\tau\simeq 10^{3}\s$
suggested by some observations,  we find $\lambda^{-1}\sim 10^{2}\s$,
or an inter-clump average distance $10^{9}-10^{10}\cm$.
The average masses of the clumps are in the order of 
$10^{18}-10^{21}\g$,
in broad agreement with the estimates found in the existing literature 
(e.g. \citealp{Ducci:2009a}).
In at least two of the SFXTs we analysed,
we were able to reproduce the flares' luminosity distributions without
invoking an accretion  barrier.

The reason why SFXT show a much higher dynamic range than
the ordinary HMXBs is still unclear:  according to our model, 
the difference is entirely due to the different properties of
the streams of matter accreting the compact objects hosted in  those systems.  HMXBs accrete from clumpy winds with a narrow mass distribution, while SFXTs from winds with an extremely wide clumps' mass  distribution. 
In a recent  paper \citet{Oskinova:2012a} pointed out 
that  a stochastic component in the clumps' velocities 
is probably essential in  shaping the light curves of SFXTs. 
In the present paper we have neglected the possible
contributions to $L(t)$ from a stochastic component of 
the velocity of the accreting stream, but it must clearly be 
 included in any  (more) realistic model.


\appendix

\section{Derivation of the generalized Fokker--Planck Equation}
\label{a-gfpe}

In this section we closely follow \citet{Denisov:2009a} to 
derive the generalized Fokker--Planck equation~\eqref{e-gfpe}
from the Langevin Equation~\eqref{e-sde-1}.
According to the Ito interpretation, the solution of Eq.~\eqref{e-sde-1}
in the (short) time interval $\Delta t$ is
\be
\label{e-ito}
\mdot(t + \Delta\,t) = \mdot(t) - \Delta \,t\, \mdot(t) /\tau
+ \delta M_{c}(t)/\tau, 
\ee
where
\be
\delta M_{c}(t)  = \int_{t}^{t + \Delta\,t}  d t' 
\:  \mdot_{c}(t') = \sum_{k=1}^{n(\Delta t)} m_{k}, 
\ee
being $n(\Delta t)$ is the number of clumps accreted in the 
time interval $\Delta t$.

In order to derive the generalised Fokker--Planck equation we introduce
the probability $p(\Delta M, \Delta t)$ that the
mass $\Delta M$ 
is accreted by the compact object in the interval
$\Delta t$
\be
\label{e-smallp}
p(\Delta M, \Delta t) = \mean{ \delta(\Delta M - \delta M_{c}(t)) }.
\ee
From this definition and Eq.~\eqref{e-smallp}, after some algebra it 
can be shown that
\bs
\begin{align}
&p(\Delta M, \Delta t) = P_{0}(\Delta t) \; \delta(\Delta M) + W(\Delta M, \Delta t)
\intertext{where}
&W(\Delta M, \Delta t) =
\sum_{k=1}^{\infty} P_{k}(\Delta t) \; 
\int d m_{1} \; \varphi(m_{1}) 
\int d m_{2}  \;\varphi(m_{2})
\cdots
\int d m_{k}  \; \varphi(m_{k})
\;
\delta\left(\Delta M - \sum_{j=1}^{k} m_{j}\right), 
\end{align}
\es
being $P_{n}(\Delta t)$ the Poisson probability  that $n$ clumps are accreted in
the time interval $\Delta t$,  and $\varphi(m)$ is the mass distribution of
the clumps. If $\Delta t$ is small, we may expand $p$ to first order 
in $\Delta t$ and take
\be
\label{e-smallp1}
p(\Delta M, \Delta t) = (1-\lambda \Delta t) \; \delta(\Delta M) 
+ \lambda \Delta t  \; \varphi(\Delta M) + {\mathcal O}(\Delta t)
\ee

The   accretion probability $P(\mdot, t)$ is defined as 
\be
P(\mdot, t) = \langle \delta(\mdot - \mdot(t)\rangle.
\ee

In order to proceed, we need a couple of formulae to  express the averages
of the functions  $\mean{F(\mdot(t))}$ and  $\mean{F(\mdot(t), \delta M_{c}(t))}$ 
in terms of the  accretion probability distributions $P(\mdot, t)$ and
$p(\Delta M, \Delta t)$.
Since the random variable $\mdot(t)$ and $\delta M_{c}(t)$ are independent, 
then 
\bs
\begin{align}
& \mean{F(\mdot(t))} = \int d \mdot \; P(\mdot, t) \; F(\mdot)
\intertext{and}
& \mean{F(\mdot(t), \delta M_{c}(t))} = 
\int d\mdot \; \int d (\Delta M)  \; 
P(\mdot, t) \;  p(\Delta M, \Delta t) \; F(\mdot, \Delta M).
\end{align}
\es
We now introduce the time Fourier transformation to derive  the generalised Fokker--Planck equation.
From the definition it is clear that
\be
P_{k}(t) = \int d \mdot \;  P(\mdot, t)  \; {\rm e}^{-i k \mdot}
= \mean{ {\rm e}^{-i k \mdot}}.
\ee
We calculate the increment
$\delta P_{k}(t) = P_{k}(t+\Delta t) - P_{k}(t)$  from Eq.~\eqref{e-ito}
as
\be
\delta P_{k}(t) = 
\left\langle 
{\rm e}^{-i k \mdot(t+\Delta t)} -{\rm e}^{-i k \mdot(t)}
\right\rangle
\ee
Expressing $\mdot(t+\Delta t)$ with Eq.~\eqref{e-ito} and
retaining only the terms linear in $\Delta t$,
\be
\delta P_{k}(t) = 
\left\langle 
{\rm e}^{-i k \mdot(t)}\left[{\rm e}^{-i k \delta\mdot_{c}(t)/\tau} -1\right]
\right\rangle
+ i k \frac{\Delta t}{\tau}
\left\langle
\mdot(t) \: {\rm e}^{-i k \mdot(t)}
\right\rangle +  {\mathcal O}(\Delta t)
\ee
Introducing the probability distributions $P$ and $p$,
\begin{eqnarray}
\delta P_{k}(t) &=& 
\int  d \mdot \; P(\mdot, t) {\rm e}^{-i k \mdot}
\int d (\Delta M ) p(\Delta M, \Delta t) 
\left[{\rm e}^{-i k \delta\mdot_{c}(t)/\tau} -1\right] +\\
&& + i k \frac{\Delta t}{\tau}
\int d \mdot P(\mdot, t) \mdot \: {\rm e}^{-i k \mdot}
+  {\mathcal O}(\Delta t).
\end{eqnarray}
Plugging Eq.~\eqref{e-smallp1} into this expression we obtain
\begin{eqnarray}
\delta P_{k}(t) &=& 
\int  d \mdot \; P(\mdot, t) {\rm e}^{-i k \mdot}
\lambda\, \Delta t \:
\int d (\Delta M ) \varphi(\Delta M)
\left[{\rm e}^{-i k \delta\mdot_{c}(t)/\tau} -1\right] +\\
&& + i k \frac{\Delta t}{\tau}
\int d \mdot P(\mdot, t) \mdot \: {\rm e}^{-i k \mdot}
+  {\mathcal O}(\Delta t) 
\end{eqnarray}
Dividing by $\Delta t$ and sending $\Delta t \to 0$
\be
\frac{\partial P_{k}}{\partial t} = 
\lambda P_{k}\; \varphi_{k/\tau}  - \lambda \: P_{k}
+ i \frac{k}{\tau}
\int d \mdot P(\mdot, t) \mdot \: {\rm e}^{-i k \mdot}.
\ee
Finally, the inverse Fourier transform yields
\be
\frac{\partial P}{\partial t} = 
- \lambda P\;  + 
\lambda \int_{0}^{\mdot\, \tau} d m  \: \varphi(m) \; P\left(\mdot - m/\tau, t \right)  
+ \frac{1}{\tau} \; \frac{\partial}{\partial \mdot}
\left(
\mdot \, P
\right), 
\ee
which is the desired Fokker--Planck equation.


\begin{acknowledgements}
This work was supported by the grant from PRIN-INAF 2009, 
"The transient X-ray sky: new classes of X-ray binaries containing neutron stars" (PI: Sidoli). 
We thank an anonymous reviewer whose comments helped us to improve
the paper.
\end{acknowledgements}


\clearpage

\end{document}